\documentclass[amsmath,amssymb,superscriptaddress,reprint, preprintnumbers,nofootinbib,notitlepage, groupdaddress,showabstract,aps,prl,twocolumn]{revtex4}
\pdfoutput=1 
             
\usepackage[utf8]{inputenc}
\usepackage[english]{babel}
\usepackage{bm}
\usepackage{graphicx}
\usepackage{dcolumn}
\usepackage{pbox}
\usepackage{comment}
\usepackage{amsmath}
\usepackage{epsfig}
\usepackage[dvipsnames]{xcolor}
\usepackage{slashed}
\usepackage{amssymb}
\usepackage{ mathrsfs }
\usepackage{color}
\usepackage{url}
\definecolor{DarkBlue}{rgb}{0.7, 0.4, 1} 
\definecolor{Blue}{rgb}{0, 0.8, 0} 
\definecolor{MyLightBlue}{rgb}{0.5,0.7,1.9}
\definecolor{MyGreen}{rgb}{0.0,0.2, 0.0}
\definecolor{MyBrickRed}{rgb}{0, 0.5, 0.2}
\RequirePackage{hyperref}
\hypersetup{colorlinks, citecolor=Blue,linkcolor=DarkBlue, urlcolor=Green}
\usepackage[normalem]{ulem}
\DeclareUnicodeCharacter{2032}{\ensuremath{'} }

\newcommand{\bea}{\begin{eqnarray}}
\newcommand{\eea}{\end{eqnarray}}

\DeclareUnicodeCharacter{2212}{-}
\newcommand{\Mzp}{M_{Z^\prime}}

\newcommand{\mdm}{m_{\rm DM}}

\begin{document}
\title{Indirect dark matter searches with neutrino telescopes via energetic cosmic showers } 
\author{Arindam Basu}
\email{arindam.basu.phy@gmail.com}
\affiliation{Department of Physics, School of Engineering and Sciences, SRM University-AP, Amaravati 522240, India}
\affiliation{School of Physical Sciences, Indian Association for the Cultivation of Science, Jadavpur, Kolkata 700032, India}
\author{Basabendu Barman}
\email{basabendu.b@srmap.edu.in}
\affiliation{Department of Physics, School of Engineering and Sciences, SRM University-AP, Amaravati 522240, India}
\author{Arindam Das}
\email{arindamdas@oia.hokudai.ac.jp}
\affiliation{Institute for the Advancement of Higher Education, Hokkaido University, Sapporo 060-0817, Japan}
\affiliation{Department of Physics, Hokkaido University, Sapporo 060-0810, Japan}
\begin{abstract}   
We explore the possibility that the high energy neutrino flux observed by terrestrial telescopes originates from dark matter (DM) annihilation. Specifically, we study a minimal, UV-complete $U(1)$ extension of the Standard Model with a Dirac DM candidate, whose annihilation into neutrinos proceeds exclusively through a $Z^\prime$ boson. {\color{black}By computing the annihilation cross section and comparing with the observed flux, we derive bounds on the model parameters.} Additional constraints are obtained within the freeze-in framework, where the observed relic
abundance is reproduced. Considering cosmic string vibrations as a source of gravitational waves, we further constrain the vacuum expectation value of the $U(1)$ breaking. All results are contrasted with perturbativity limits and existing constraints from low- and high-energy experiments.
\end{abstract} 
\maketitle
\section{Introduction}
The absence of any significant signal in collider, direct and indirect searches, has placed strong constraints on the standard weakly interacting massive particle (WIMP) dark matter (DM) paradigm~\cite{Jungman:1995df,Roszkowski:2017nbc,Arcadi:2017kky,Arcadi:2024ukq}, in which the relic abundance arises through thermal freeze-out, typically with a weak-scale interaction cross-section. As an alternative, several mechanisms of DM production have been proposed, among which freeze-in has attracted considerable attention in recent years~\cite{Hall:2009bx,Bernal:2017kxu}. The key feature of this framework is the extremely feeble interaction between DM and the Standard Model (SM), which is necessary to ensure non-thermal, out-of-equilibrium production. While this tiny coupling makes experimental detection highly challenging, recent studies have shown that it still can be probed in scenarios with modified cosmological histories~\cite{Allahverdi:2020bys}, and potentially at intensity/energy-frontier experiments~\cite{Belanger:2018sti,Okada:2020cue,Ghosh:2024nkj,Barman:2024tjt}, gravitational wave detectors~\cite{Bian:2018mkl,Bian:2018bxr,Shibuya:2022xkj} or even direct-detection facilities~\cite{Hambye:2018dpi,Duch:2017khv,Bhattiprolu:2022sdd,PhysRevD.111.063537,PhysRevLett.130.031803,Das:2023cbv,Arcadi:2024obp}. 

{\color{black} In recent years, astroparticle observations have witnessed remarkable progress. The IceCube Collaboration has reported a $4.2\sigma$ excess of neutrinos with energies in the range $1.5$--$15\,\text{TeV}$, spatially associated with the Seyfert galaxy NGC~1068~\cite{IceCube:2022der}. This observation has attracted considerable attention because no corresponding high-energy photon flux has been detected at the level expected from conventional hadronic production mechanisms. IceCube has also observed a $\sim 290\,\text{TeV}$ neutrino event (IC-170922A) in coincidence with $\gamma$-ray emission from the blazar TXS~0506+056~\cite{IceCube:2018dnn,IceCube:2022der}, pointing to the existence of powerful cosmic accelerators capable of producing extremely energetic neutrinos. Alternatively, such high-energy neutrinos could originate from DM annihilation\footnote{A comprehensive study of DM annihilation directly to neutrino-antineutrino pairs have been presented in~\cite{Arguelles:2019ouk}, for  DM over a mass range of 1 MeV to $10^{12}$ GeV.} within dense DM spikes surrounding supermassive black holes~\cite{Gondolo:1999ef,Shapiro:2022prq}. Complementary observations have been performed by other neutrino telescopes. The Astronomy with a Neutrino Telescope and Abyss environmental RESearch (ANTARES) detector collected data on $\mathcal{O}(20\,\text{TeV})$ neutrinos over a period of fifteen years (2007--2022) in the Mediterranean Sea~\cite{ANTARES:2024ihw}, while earlier ANTARES data were used to place constraints on the thermally averaged DM annihilation cross section~\cite{ANTARES:2019svn}. Similarly, the Baikal Gigaton Volume Detector (Baikal-GVD) reported the observation of a diffuse cosmic neutrino flux during 2018--2021~\cite{Baikal-GVD:2018isr}, recording a significant excess of cascade-like events over the expected atmospheric background, consistent with IceCube observations~\cite{Baikal-GVD:2022fis}. More recently, the KM3NeT Collaboration reported the detection of event KM3-230213A~\cite{KM3NeT:2025npi}, corresponding to an ultra-high-energy neutrino with energy in the range $110\,\text{PeV}\leq E_\nu\leq 790\,\text{PeV}$ and a median energy of $220\,\text{PeV}$, making it the most energetic neutrino ever detected on Earth. Within standard astrophysical scenarios, such extreme-energy neutrinos are expected to be produced through proton--proton or proton--photon interactions and should therefore be accompanied by high-energy photon emission. However, dedicated multi-messenger follow-up studies have not identified any convincing electromagnetic counterpart~\cite{KM3NeT:2025bxl,KM3NeT:2025aps,KM3NeT:2025vut}, raising the possibility of a cosmogenic origin.

Nevertheless, the inferred flux associated with KM3-230213A carries an uncertainty of order $\mathcal{O}(3)$, significantly exceeding the predictions of current ultra-high-energy neutrino and $\gamma$-ray models, as well as the sensitivities of IceCube and the Pierre Auger Observatory (PAO)~\cite{PierreAuger:2022atd}. Furthermore, neither experiment has reported a comparable event at a significance level of $2.5\sigma$--$3\sigma$. This apparent tension~\cite{Li:2025tqf,KM3NeT:2025ccp,Palmisano:2025abd} challenges a purely cosmogenic interpretation and motivates the exploration of alternative scenarios, including DM annihilation in energetic astrophysical environments, as a possible source of these high-energy neutrino events.
}

Alongside these neutrino observations, current detections of Gravitational Waves 
(GWs)~\cite{LIGOScientific:2021nrg,Caldwell:2022qsj,NANOGrav:2023gor,NANOGrav:2023hvm,Xu:2023wog,EPTA:2023fyk,LISACosmologyWorkingGroup:2022jok} 
have opened a complementary avenue to probe BSM physics. For example, if symmetries are broken 
spontaneously at very high temperatures, topological defects such as cosmic strings and domain walls 
may form in the early Universe~\cite{Nielsen:1973cs,Kibble:1976sj}, and these defect networks can act 
as prominent sources of a stochastic GW background, with the associated symmetry-breaking scale linked to new physics.

In this work, we investigate high energy cosmic showers from active galaxies and blazars, considered potential sources of ultra-high-energy (UHE) neutrinos observed at IceCube, ANTARES, Baikal-GVD, KM3NeT, and PAO. We propose that such neutrinos may originate from DM annihilation in these astrophysical environments\footnote{New physics interpretation of KM3NeT events, based on transient astrophysical sources of DM has been explored, for example, in Refs.~\cite{Farzan:2025ydi,Dev:2025czz}.}, producing jets subsequently detected on Earth. As a concrete realization, we focus on a UV-complete, minimal extension of the SM with an anomaly-free $U(1)_X$ gauge symmetry~\cite{Oda:2015gna,Das:2016zue}, supplemented by three generations of SM-singlet right-handed neutrinos (RHNs) and a singlet scalar field whose nonzero vacuum expectation value (VEV) breaks the $U(1)_X$\footnote{Bounds on the parameter space of $\sim\mathcal{O}\left(\text{TeV}-\text{PeV}\right)$ DM annihilating into long-lived dark gauge boson mediator, that further decay into neutrinos has been obtained in~\cite{Nguyen:2022zwb}.}. This breaking generates Majorana masses for RHNs, which induce small neutrino masses and flavor mixing via the seesaw mechanism~\cite{Minkowski:1977sc,Yanagida:1979as,Gell-Mann:1979vob,Mohapatra:1979ia,Schechter:1980gr}. Within this framework, we introduce a Dirac DM candidate annihilating via the $U(1)_X$ gauge boson $Z^\prime$, producing energetic neutrino pairs. By comparing with current neutrino flux data, we derive bounds on the $[g_X,\, \Mzp]$ parameter space in a UV complete model framework. We also explore the feeble DM--SM interactions required for freeze-in production, mapping the relic density consistent parameter space. Finally, we highlight complementary probes from upcoming GW detectors, future IceCube-Gen2, existing beam-dump experiments, low-energy scattering, and collider searches.
\\



\section{The framework}
General $U(1)$ extension of the SM can be realized as SM $\otimes\,U(1)_X$ scenario where $U(1)_X$ gauge group is considered as the linear combination of SM $U(1)_{\rm{Y}}$ and the $U(1)_{\rm{B-L}}$ gauge groups that are anomaly free. As a result, we can write the $U(1)_X$ charges $(Q_X)$ of the particles as $Q_X= x_H Q_{\rm{Y}}+ x_\Phi Q_{\rm{B-L}}$ where $Q_Y$ represents the SM $U(1)_{\rm{Y}}$ hyper-change and $Q_{\rm{B-L}}$ stands for the B$-$L (baryon minus lepton) charge, respectively. In this scenario three generations of the SM quarks transforms as $q_L^i=\{3,2,\frac{1}{6}, x_q=\frac{1}{6} x_H+\frac{1}{3} x_\Phi\}$, $u_R=\{3,1,\frac{2}{3}, x_u=\frac{2}{3} x_H+\frac{1}{3} x_\Phi\}$ and $d_R^i=\{3,1,-\frac{1}{3}, x_d=-\frac{1}{3}x_H+\frac{1}{3} x_\Phi\}$, respectively. Three generations of SM leptons transform as $\ell_L^i=\{1,2,-\frac{1}{2}, x_\ell=-\frac{1}{2}x_H-x_\Phi\}$ and $e_R^i=\{1,1,-1, x_e=-x_H-x_\Phi\}$, respectively  where $i$ stands for the three generations of these fermions. The SM-like Higgs field transforms following $H=\{1,2,\frac{1}{2}, x_h=\frac{1}{2}x_H\}$ and an SM-singlet scalar, responsible for the $U(1)_X$ breaking, transforms following $\Phi=\{1,1,0, x_\Phi\}$. There are three generations of SM-singlet RHNs follow the transformation rule $N_R^{i}=\{1,1,0, -x_\Phi\}$. These RHNs are required to generate light neutrino mass and flavor mixing through the seesaw mechanism \cite{Oda:2015gna,Das:2016zue}. In our analysis we fix $x_\Phi=1$ without the loss of generality. Considering $x_H=0$ the scenario reduces to the B$-$L model. In addition to that if we consider $x_H=-1 (1)$, $U(1)_X$ charge of $e_R (d_R)$ will vanish. For $x_H=2$, $U(1)_X$ charge of left and right handed fermions will not vanish creating a purely chiral scenario. The relevant for the Lagrangian required for our work manifesting the light neutrino mass can be written as 
\bea
\mathcal{L}_{\rm{yuk}}\supset -Y_{\nu_{ij}} \overline{\ell_L^i} \widetilde{H} N_R^j -\frac{1}{2} Y_{N_\alpha} \overline{\left({N_R^\alpha}\right)^c}\,N_R^\alpha\,\Phi+\text{h.c}\,,
\label{eq:LYk}
\eea
with $\widetilde{H}=i\sigma^2\,H^\star$, where $\sigma^2$ is the Pauli matrix. The renormalizable scalar potential involving $H$ and $\Phi$ fields can be written as 
\bea
V=\sum_{\mathcal{I}= H, \Phi} \Big[m_{\mathcal{I}}^2 (\mathcal{I}^{\dagger} \mathcal{I})+ \lambda_{\mathcal{I}} (\mathcal{I}^\dagger \mathcal{I})^2 \Big] +
\lambda^\prime (H^\dagger H) (\Phi^{\dagger} \Phi)\,.
\label{pot}
\eea 
After the breaking of $U(1)_X$ and electroweak gauge symmetries, the scalar fields $H$ and $\Phi$ develop respective VEVs following
\begin{align}\label{eq:VEV}
  \langle H\rangle \ = \ \frac{1}{\sqrt{2}}\begin{pmatrix} v+h\\0 
  \end{pmatrix}~, \quad {\rm and}\quad 
  \langle\Phi\rangle \ =\  \frac{v_\Phi^{}+\phi}{\sqrt{2}}~,
\end{align}
where the electroweak scale is $v=246$ GeV at the potential minimum. Taking $v_\Phi\gg v$, in our consideration, we obtain the mass of the $U(1)_X$ gauge boson as $M_{Z^\prime}^{}= 2 g_X  v_\Phi$. We consider negligible mixing\footnote{Current bounds from LHC~\cite{Robens:2015gla,Chalons:2016jeu,Das:2022oyx}, LEP~\cite{LEPWorkingGroupforHiggsbosonsearches:2003ing}, prospective colliders like ILC~\cite{Wang:2020lkq} and CLIC~\cite{CLIC:2018fvx} suggests that the scalar mixing $<0.001$, for $\phi$ mass up to 1 TeV.} between the physical states of $H$ and $\Phi$, as well as between $Z$ and $Z^\prime$ for simplicity\footnote{LEP data places an upper bound on the kinetic mixing $<10^{-2}$~\cite{Leike:1992uf,Carena:2004xs}, rendering kinetic-mixing--induced scattering processes inefficient.}. After the breaking of $U(1)_X$ symmetry, the Majorana mass of the RHNs $M_{\alpha}= Y_{N_\alpha} v_\Phi/\sqrt{2}$ and after the electroweak symmetry the Dirac mass of the light left-handed neutrinos $m_{{D}_{\alpha \beta}}= Y_{{\nu}_{\alpha \beta}}v /\sqrt{2}$ are generated from Eq.~\eqref{eq:LYk}, respectively. Hence using the well known seesaw formula mass of the light active neutrinos can be generated as $-m_D^{} M_\alpha^{-1} m_D^T$~\cite{Gell-Mann:1979vob,Sawada:1979dis, Mohapatra:1980yp} which successfully explains the origin of tiny neutrino mass and flavor mixing. 

We incorporate a SM gauge singlet fermion $\chi_{L,R}$, which is considered to be a potential DM candidate with $U(1)_X$ charge $n_\chi$. As a result, it interacts only with $Z^\prime$ via 
\bea
\mathcal{L}_{\rm{kin}} = i\,\overline{\chi_L}\,\gamma_\mu \mathcal{D}^\mu \chi_L - i\,\overline{\chi_R}\,\gamma_\mu \mathcal{D}^\mu \chi_R\,,
\label{Xm}
\eea
where $\mathcal{D}^\mu= \partial^\mu + i g_X n_\chi {Z^\mu}^\prime$. We forbid the choices of $n_\chi=\pm x_\Phi, \pm 3 x_\Phi$ to 
prevent the decays of $\chi_{L, R}$ into the RHNs and other particles ensuring stability of the DM. The interaction Lagrangian of the DM candidate is given by,
\bea
\mathcal{L}_{\rm{DM}} = i \overline{\chi} \gamma_\mu \partial^\mu \chi- g_X n_\chi \gamma_\mu {Z^\mu}^\prime \chi + (m_\chi \overline{\chi_L} \chi_R + \text{h.c.})\,,
\label{XDM}
\eea
where $\chi= \chi_L+\chi_R$, and we have added a Dirac mass term for the DM. We consider an aspect which is sensitive to the UV theory. Hence through non-renormalizable and higher dimensional operators neutrinos could mix with the DM candidate for odd $n_\chi$. Therefore we restrict ourselves choosing $n_\chi$ to be even and fractional numbers. In the entire analysis we choose $n_\chi=100$ and $10000$ and $m_\chi=3 M_{Z^\prime}$ to prohibit the decay of $Z^\prime$ into DM. {\color{black} It is worth mentioning here, the benchmark values $n_\chi=100$ and $10000$ are chosen to explore the phenomenological consequences of an enhanced dark-sector coupling $g_\chi \equiv n_\chi g_X$. Since the relevant interactions depend only on the combination $g_\chi$, these large charge assignments should be viewed as an effective parametrization of strong DM--$Z'$ interactions that can improve the detection aspect of freeze-in DM, rather than as a statement about a specific UV-complete realization. Throughout our analysis we impose the perturbativity\footnote{{\color{black} Perturbative unitarity constraints on the B$-$L model, derived using partial-wave analysis, have been studied, for example, in Refs.~\cite{Basso:2010jt,Basso:2010jm,Basso:2010hk,Basso:2011na}. We find that the bounds obtained from requiring perturbativity up to the Planck scale are weaker than existing collider constraints and are therefore not shown.}} requirement $n_\chi\,g_X<\sqrt{4\pi}$. }
\\

\section{DM-annihilation through $Z^\prime$ from cosmic events}
We model the DM halo density using the Navarro--Frenk--White (NFW) profile~\cite{Navarro:1995iw,Navarro:1996gj},  
\begin{align}
    \rho_{\text{NFW}}(r) = \rho_s \left(\frac{r}{r_s}\right)^{-1}
    \left(1+\frac{r}{r_s}\right)^{-2}\,,
\end{align}
where $r_s$ is the scale radius and $\rho_s$ the associated scale density. {\color{black} The adiabatic growth of a supermassive black hole (SMBH) at the Galactic center, together with the potential self-annihilation of DM particles, alters this distribution and gives rise to a central density spike. If the surrounding profile remains unaffected, the resulting DM distribution can be expressed as~\cite{Quinlan:1994ed,Gondolo:1999ef},
\begin{align} \label{eq:DM-profileNGC}
    \rho_\chi(r) = 
    \begin{cases}
        0, & r < 4R_\text{S}\,, \\[10pt]
        \dfrac{\rho_{\text{sp}}^{7/3}(r)\,\rho_c}
              {\rho_{\text{sp}}^{7/3}(r)+\rho_c}, 
              & 4R_\text{S} \leq r \leq R_{\text{sp}}\,, \\[10pt]
        \dfrac{\rho_{\text{NFW}}(r)\,\rho_c}
              {\rho_{\text{NFW}}(r)+\rho_c}, 
              & r \geq R_{\text{sp}}\,,
    \end{cases}
\end{align}
For the DM distribution at the center of the Milky Way (MW), taking into account significant stellar heating  due to the known existence of S-stars that orbit extremely close to the supermassive black hole Sgr $\text{A}^\star$~\cite{Sadeghian:2013laa,Balaji:2023hmy,Akita:2025dhg},
\begin{align} \label{eq:DM-profileMW}
    \rho_\chi^{\text{MW}}(r) = 
    \begin{cases}
        0, & r < 4R_\text{S}^{\text{MW}}\,, \\[10pt]
        \dfrac{\rho_{\text{sp}}^{3/2}(r)\,\rho_c}
              {\rho_{\text{sp}}^{3/2}(r)+\rho_c}, 
              & 4R_\text{S}^{\text{MW}} \leq r \leq R_{\text{sp}}^{\text{MW}}\,, \\[10pt]
        \dfrac{\rho_{\text{NFW}}(r)\,\rho_c}
              {\rho_{\text{NFW}}(r)+\rho_c}, 
              & r \geq R_{\text{sp}}^{\text{MW}}\,,
    \end{cases}
\end{align}
}
where $R_{\text{S}} = 2GM_{\text{BH}}$ is the Schwarzschild radius of a black hole of mass $M_{\text{BH}}$, and $R_{\text{sp}}$ denotes the spike radius. {\color{black}Similarly, $R_{\rm S}^{\rm MW}$ and $R_{\rm sp}^{\rm MW}$ denote the Schwarzschild radius and spike radius, respectively, for the MW.} The spike profile in the absence of DM annihilation is given by
\begin{align}
    \rho_{\text{sp}}^\gamma(r) 
    = \rho_{\text{R}} 
      \left(1-\frac{4R_\text{S}}{r}\right)^3 
      \left(\frac{R_{\text{sp}}}{r}\right)^\gamma\,,
\end{align}
with $\rho_{\text{R}} = \rho_s\,(r_s/R_{\rm sp})$ chosen to ensure continuity with the NFW profile at $r=R_{\text{sp}}$. Including DM annihilation, the spike reaches a maximum density  
\begin{align}
    \rho_c = \frac{m_\chi}{\langle \sigma v\rangle\,t_{\text{BH}}}\,,
\end{align}
where $m_\chi$ is the DM mass, $t_{\rm BH}$ is the age of the SMBH. {\color{black}Considering NGC 1068 or such an extragalactic source does not contain stars close to its central SMBH, we follow the DM profile described as Eq.~\ref{eq:DM-profileNGC}, while for MW, we follow Eq.~\ref{eq:DM-profileMW}. For NGC 1068, we adopt the benchmark values $R_{\text{sp}} = 0.7$ kpc, $M_{\text{BH}} = 10^7 M_{\odot}$, $\rho_{s} = 0.35\,\text{GeV}\,\text{cm}^{-3}$,
$\gamma = 7/3$, $r_s = 13$ kpc, and $t_{\text{BH}} = 10^9$ years~\cite{Bahcall:1976aa,KA:2023dyz}. Whereas for MW, we take $R_{\text{sp}}^{\text{MW}} = 0.34$ pc, $M_{\text{BH}} = 4.3 \times 10^{6} M_{\odot}$, $\rho_{s}^{\text{MW}} = 0.351\,\text{GeV}\,\text{cm}^{-3}$, $\gamma = 3/2$, $r_s^{\text{MW}} = 18.6$ kpc, and $t_{\text{BH}} = 10^{10}$ years~\cite{Balaji:2023hmy}.}
Conservatively assuming the point of interest is the Galactic center, we obtain the neutrino flux from the DM annihilation within the halo, over the solid angle $\Delta \Omega$,
\begin{align}\label{eq:nu-flux1}
    \frac{d\phi_\nu}{dE_\nu} = \frac{\langle \sigma v \rangle}{8\pi\, m_\chi^2} \frac{1}{3} \frac{dN_\nu}{dE_\nu} \int_{\Delta \Omega} d\Omega\int_0^{R_{\text{max}}} ds \, \rho_\chi^2(r)\,.
\end{align}
{\color{black}Here, the DM density profile $(\rho_{\chi}(r))$ follows Eq.~\ref{eq:DM-profileNGC} for the extragalactic evolution, whereas Eq.~\ref{eq:DM-profileMW} is adopted for the flux calculation of the MW event.} The thermally averaged DM annihilation cross section into leptonic final states is denoted by $\langle \sigma v\rangle$, given as,
\bea
\bigl<\sigma v \bigr>_{\ell\bar{\ell}}=
\frac{n_\chi^2 g_X^4 }{ 2\pi m_\chi^2 }\,\frac{ \sqrt{ 1 - x_\ell^2 }\,(x_H/2+x_\Phi^{})^2 ( 1 - x_\ell^2)}{(x_{z^\prime}^2-4)^2 + x_{z^\prime}^4\Gamma_{Z^\prime}^2/M_{Z^\prime}^2}\,, 
\eea
where $x_{\ell(z^\prime)}^{}=m_\ell^{}\,(\Mzp)/m_\chi^{}$ and $\Gamma_{Z^\prime}$ is the total decay width of $Z^\prime$. {\color{black} The factor of $1/3$ in Eq.~\eqref{eq:nu-flux1} accounts for flavor averaging at Earth. Since the neutrinos originate from the Galactic halo and propagate over astrophysical distances, flavor oscillations effectively decohere the neutrino states. For the flavor compositions relevant to DM annihilation into SM final states, the oscillated neutrino flux arriving at Earth is approximately distributed equally among the three flavors, $\left(\nu_e:\nu_\mu:\nu_\tau\right)_\oplus \simeq \left(1:1:1\right)$.
Therefore, the flux of a given neutrino flavor is taken to be one-third of the total neutrino flux. }

The partial decay widths of $Z^\prime$ into a pair of SM fermions can be written as,
\bea
\Gamma(Z' \to \bar{f} f)&=& N_C^{} \frac{M_{Z^\prime}^{} g_{X}^2}{24 \pi} \Bigg[ \left( q_{f_L^{}}^2 + q_{f_R^{}}^2 \right) \left( 1 - \frac{m_f^2}{M_{Z^\prime}^2} \right)+ \nonumber \\
&+&6 q_{f_L^{}}^{} q_{f_R^{}}^{} \frac{m_f^2}{M_{Z^\prime}^2}\Bigg] \left(1-\frac{4\,m_f^2}{M_{Z^\prime}^2} \right)^{\frac{1}{2}}\,,
\label{eq:width-ll}
\eea   
where $m_f^{}$ stands for the mass of the SM fermions, $q_{f_{L(R)}}$ is the $U(1)_X$ charge of the left (right) handed fermions, $N_C^{}=1 (3)$ being the color factor for the SM leptons (quarks). Here, $dN_\nu/dE_\nu$ denotes the neutrino energy spectrum, and the factor $1/3$ accounts for averaging over the three neutrino flavors. The distance from the Galactic center is parametrized as,
\begin{align}
    r^2(l,\theta) = l^2 + R_{\odot}^2 -2\,l\,R_\odot\, \cos \theta\,,
\end{align}
where $l$ is the distance along the line of sight (l.o.s) from the observer and $\theta$ is the angle between the direction of the galactic center and the line-of-sight. In the neutrino flux calculation, the l.o.s is evaluated up to $R_{\text{max}} = \sqrt{R_{\rm vir}^2 - R_\odot^2\,\sin^2 \theta} + R_\odot \cos \theta$ to account for the finite extent of the dark matter halo, with $R_{\rm vir}=200$ kpc~\cite{Balaji:2023hmy} being the halo's virial radius, and $R_\odot = 8.5$ kpc is the distance of the galactic center from the sun. The flux is evaluated for opening angle $\theta\in[0^\circ,\,10^\circ]$. We assume the neutrino generation is due the DM annihilation $\chi\, \chi \to \nu_i \nu_i$ ($i=1,2,3$ for the three flavor of neutrinos), mediated by $Z'$. Consequently, the RHS of Eq.~\eqref{eq:nu-flux1} becomes a function of $g_X$ and $\Mzp$. Thus, in the rest frame of the annihilating DM,
\begin{align}
    \frac{dN_\nu}{dE_\nu} = \delta (E_\nu - m_\chi)\,,
\end{align}
which corresponds to a monochromatic neutrino line. In practice, detector energy resolution broadens this line, smearing the $\delta$-function around the peak energy. We model the response as a Gaussian energy resolution of width $w = 0.25$ in $\text{log}_{10} E$(GeV), centered around the peak line \cite{Palomares-Ruiz:2007egs,Akita:2025dhg}. The observed flux is then obtained as,
\begin{align} \label{eq:gamma-flux3}
    \frac{d\phi_{\nu}}{dE_{\nu}} = \int_0^\infty dE' \, \frac{d\phi_{\nu}}{dE'} \mathcal{G}(E',E_\nu),
\end{align}
where 
\begin{align} \label{eq:Convolv-func}
    \mathcal{G}(E',E_\nu) = \frac{e^{-\frac{1}{2}(w\,\text{ln}\,10)^2}}{\sqrt{2\pi\,w} \,\text{ln}\,10}\, \frac{1}{E'} e^{-\frac{1}{2}\Big( \frac{\text{log}_{10}\, E_\nu/E'}{w}\Big)^2}\,,
\end{align}
is the convolution function that accounts for detector resolution. {\color{black} For a dark matter mass $m_\chi$, the neutrino spectrum at production is monochromatic, i.e. $\frac{dN_\nu}{dE_\nu}\propto \delta(E_\nu-m_\chi)$. However, the quantity compared with observations is not the injected spectrum but the detector-smeared spectrum. We convolve the monochromatic spectrum with a Gaussian energy-response function following Eq.~\eqref{eq:Convolv-func}. The neutrino telescopes considered in this work typically have energy resolutions of order (10\%) for the relevant event classes and energy ranges. Consequently, the observed spectrum is broadened into a continuous distribution around ($E_\nu=m_\chi$), allowing a direct comparison with measurements reported over finite energy bins. Therefore, although the source spectrum is monochromatic, the detector-convolved spectrum extends over a finite energy interval and can be compared with the observed flux data.}

{\color{black}
Before proceeding further it is worth mentioning here that our analysis is not based on an event-level likelihood or on a dedicated reanalysis of the experimental datasets. Instead, we perform a phenomenological flux comparison. For a given choice of $(m_\chi, M_{Z'})$, we calculate the detector-convolved neutrino flux and compare its magnitude with the reported observational flux in the corresponding energy range. We then determine the coupling $g_X$ required for the predicted flux to reach the observed level. Therefore, the results should be interpreted as ``flux-matching estimates" of the viable parameter space rather than statistically rigorous exclusion limits derived from a likelihood analysis.
}
\\

\section{Complementarity with DM freeze-in and GWs}
{\color{black} In this section, we discuss the freeze-in production of DM in our framework and show how complementary searches using GWs can be used to constrain the corresponding DM parameter space. Our analysis closely follows Refs.~\cite{Barman:2024lxy,Barman:2025bir}, where the freeze-in scenario was studied using ultra-high-energy neutrino events observed at IceCube~\cite{IceCube:2013cdw, IceCube:2013low, IceCube:2014stg,IceCube:2015gsk,IceCube:2015qii,Li:2025tqf} and KM3NeT~\cite{KM3NeT:2025npi,KM3NeT:2025bxl,KM3NeT:2025aps,KM3NeT:2025vut}. Here, we extend that approach by incorporating data from multiple neutrino telescopes, allowing us to derive stronger and more complementary constraints on the feeble DM--SM interactions responsible for generating the observed DM relic abundance through freeze-in.

It is worth noting that the DM considered here can also achieve the observed relic abundance through the conventional freeze-out mechanism via the process DM\,DM $\rightarrow$ SM\,SM mediated by the $Z'$ boson. Such $Z'$-portal Dirac DM scenarios have been extensively studied in the literature. For instance, Refs.~\cite{Mohapatra:2019ysk,Nath:2021uqb} focused on the region $\Mzp\lesssim 10$ GeV to investigate constraints from low-energy fixed-target and intensity-frontier experiments, while Ref.~\cite{Klasen:2016qux} examined the compatibility of the model with high-energy collider searches. In contrast, our primary objective is to determine how effectively existing collider and fixed-target experiments can probe the extremely small couplings required for freeze-in production, rather than the relatively large couplings, $g_X\sim\mathcal{O}(1)$, typically associated with standard freeze-out. At the same time, we explore the potential of neutrino telescope observations to constrain the freeze-in parameter space. Further, as pointed out in the beginning, we focus on the freeze-in framework because conventional freeze-out scenarios are already subject to strong constraints from DM--SM scattering experiments (direct detection). This motivates the study of whether current and future experiments at both the energy and intensity frontiers can probe DM--SM interactions with couplings as small as $\lesssim \mathcal{O}(10^{-5})$. 

Before presenting our results, we briefly review the freeze-in production mechanism and the GW spectrum generated by a cosmic string network.
}
\\

(i) {\it $Z'$-mediated Freeze-in:} The DM is produced predominantly through two processes: (a) on-shell decays of the $Z'$ boson into DM pairs, and  (b) $Z'$-mediated $s$-channel scattering of thermal bath particles into DM pairs. We forbid the decay channel, by considering $m_\chi = 3M_{Z'}$. The DM number density evolution can be tracked by solving the Boltzmann equation (BEQ), conveniently expressed in terms of the yield $Y_{\rm DM}\equiv n_{\rm DM}/s$ as,
\begin{align}\label{eq:beq}
& x\,H\,s\,\frac{dY_{\rm DM}}{dx} = \gamma_{\mathfrak{s}}\,, & x \equiv m_\chi/T\,,
\end{align}
where $\gamma_{\mathfrak{s}}$ is the reaction density for $2\to2$ processes,
\begin{align}
& \gamma_{\mathfrak{s}} = \frac{T}{32\pi^4}\,g_a g_b 
\int_{\mathfrak{s}_{\rm min}}^\infty d\mathfrak{s}\,\frac{\big[(\mathfrak{s} - m_a^2 - m_b^2)^2 - 4m_a^2 m_b^2\big]}{\sqrt{\mathfrak{s}}}
\nonumber\\&
\sigma(\mathfrak{s})_{a,b\to1,2}\,K_1\!\left(\frac{\sqrt{\mathfrak{s}}}{T}\right)\,.
\label{eq:gam-ann}
\end{align}
Here $g_{a,b}$ are the internal degrees of freedom of the initial states $a,b$, and the lower limit of the integration is $\mathfrak{s}_{\rm min} = \max\!\left[(m_a + m_b)^2,\,4m_\chi^2\right]$. During radiation domination, the Hubble rate and entropy density become~\cite{Kolb:1990vq},
\begin{align}
H(T) &= \frac{\pi}{3}\sqrt{\frac{g_*(T)}{10}}\,\frac{T^2}{M_P}\,,
\nonumber\\
s(T) &= \frac{2\pi^2}{45}\,g_{*s}(T)\,T^3\,,
\end{align}
where $g_*(T)$ and $g_{*s}(T)$ denote the relativistic degrees of freedom (DOF) contributing to the energy and entropy densities, respectively. 
\begin{figure*}
\centering    
\includegraphics[width=0.497\textwidth,angle=0]{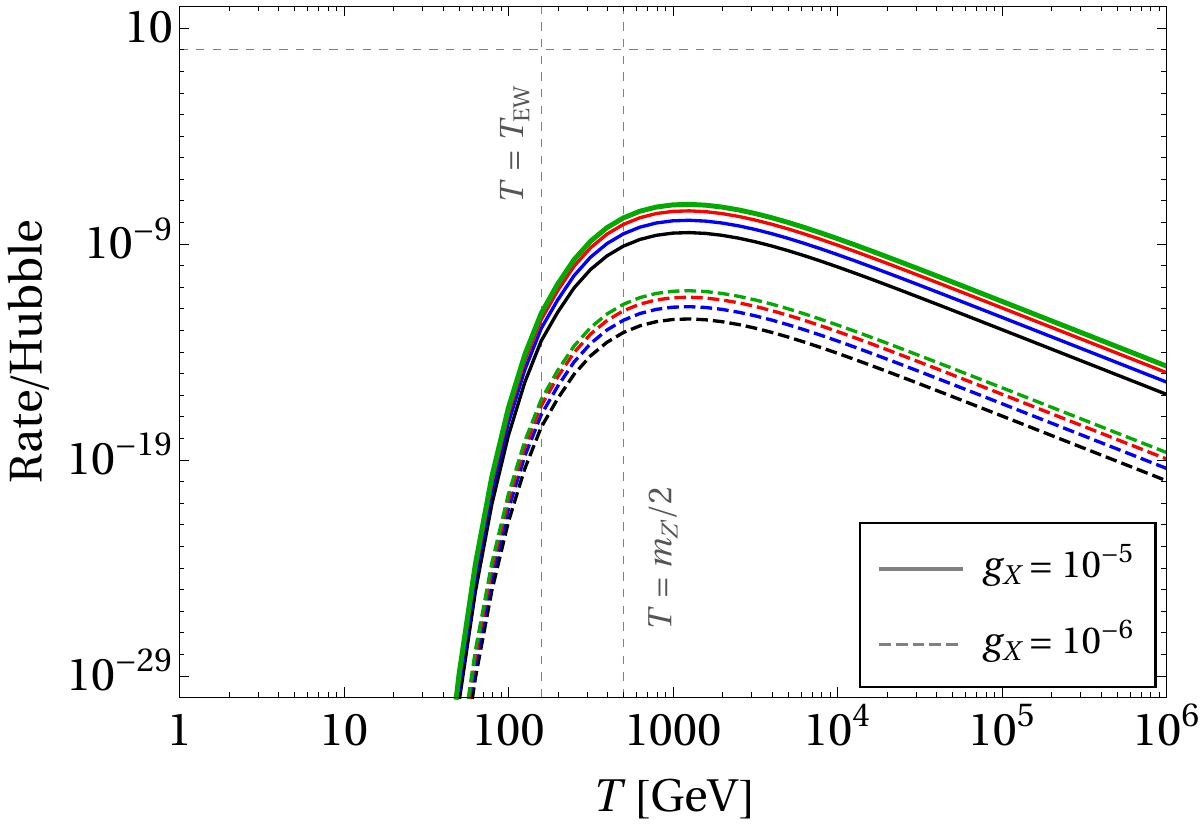}
\includegraphics[width=0.497\textwidth,angle=0]{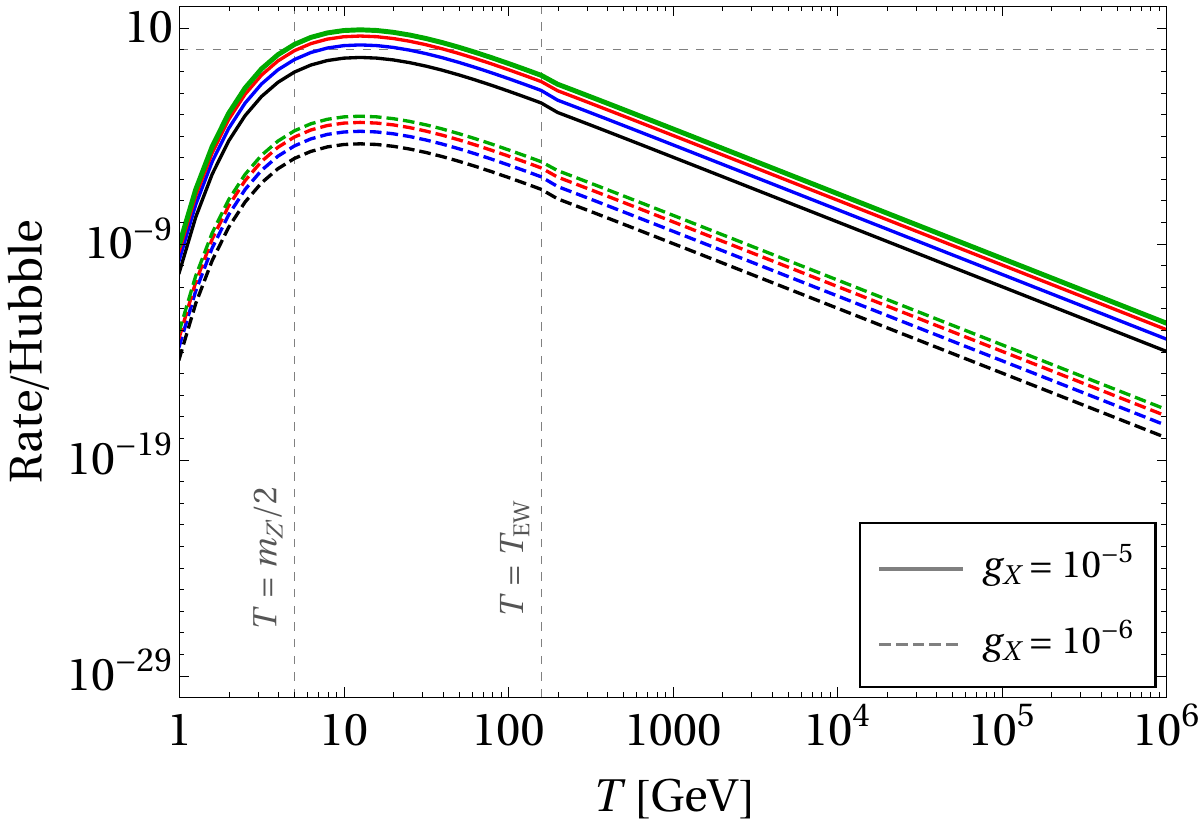}\\[10pt]
\includegraphics[width=0.497\textwidth,angle=0]{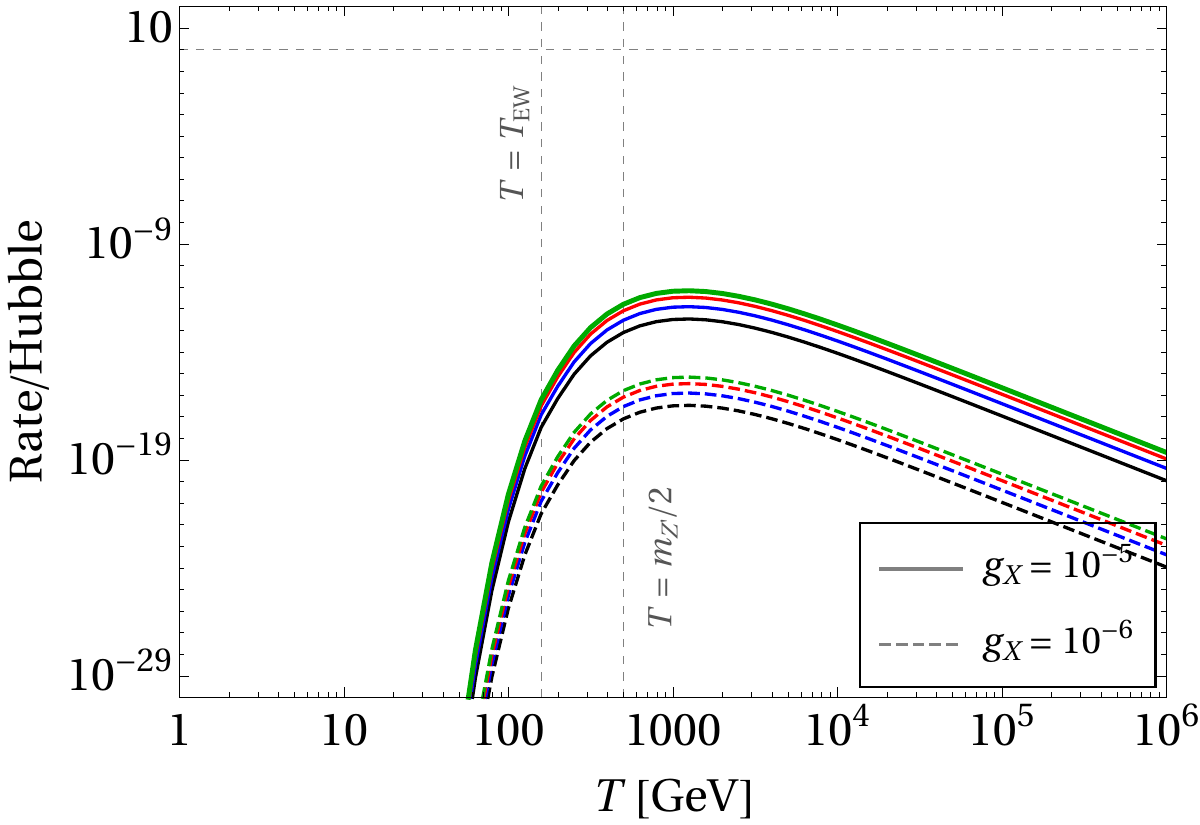}
\includegraphics[width=0.497\textwidth,angle=0]{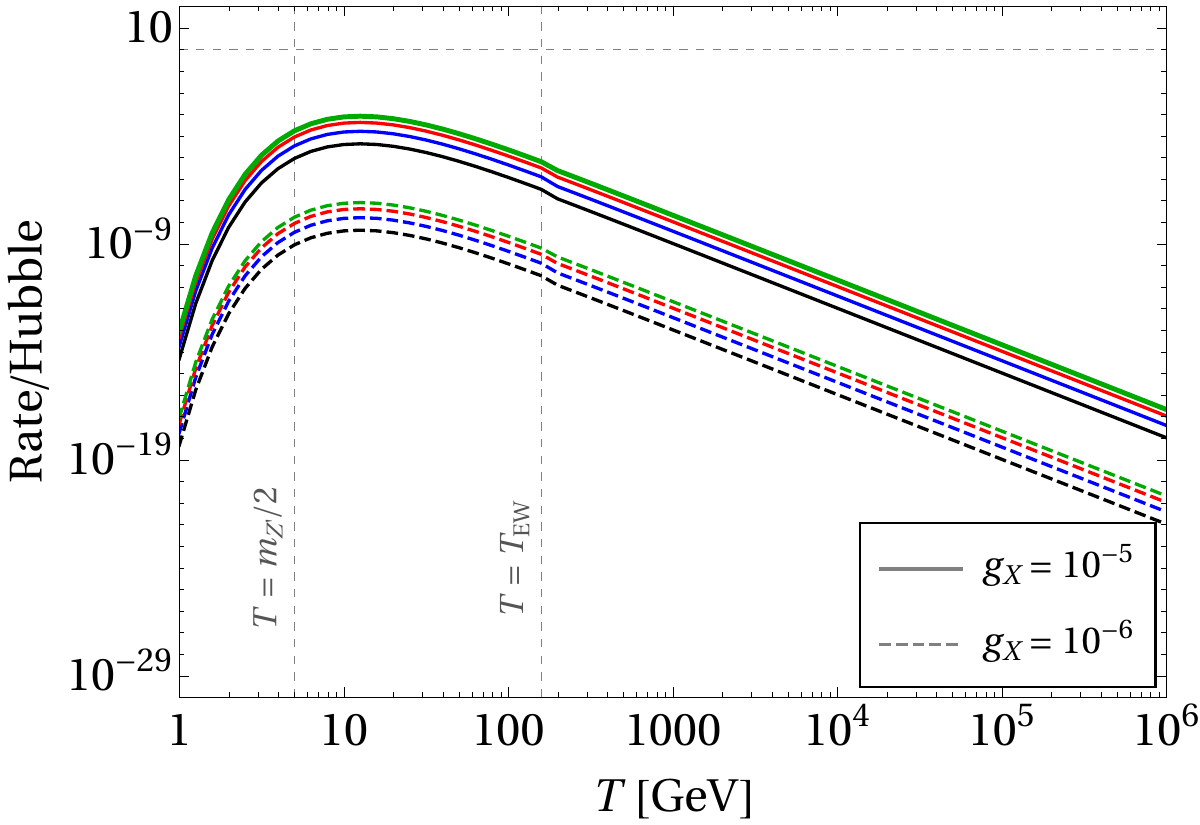}
\caption{Comparison of DM-SM interaction rate with the Hubble expansion rate, as a function of temperature for $\mdm=3\,\Mzp$, with $\Mzp=\{1\,\text{TeV},\,10\,\text{GeV}\}$ in the left and in the right panel, respectively. The red, gree, blue and black curves correspond to $x_H=\{-1,\,0,\,1,\,2\}$, respectively. The horizontal gray dashed line denotes the equity of interaction rate and Hubble rate. In top panel we have fixed $n_\chi=10^4$, while in the bottom panel we choose $n_\chi=10^2$.}
\label{fig1}
\end{figure*}
In the regime where the masses of the initial and final states are negligible compared to the center-of-mass energy of the process, the total DM production cross section takes the form  
\begin{align}
&\sigma(s) \simeq 
\frac{g_X^4\,n_\chi^2\,\left(233\,x_H^2 + 572\,x_H\,x_\Phi + 464\,x_\Phi^2\right)}{864\,\pi}
\nonumber\\&
\times\frac{s}{(s - \Mzp^2)^2 + \Gamma_{Z'}^2\,\Mzp^2}\,,
\end{align}
where we have employed the results of Refs.~\cite{Barman:2024lxy,Barman:2025bir}. {\color{black} From this, one obtains an approximate analytical expression for the reaction density $\gamma_\mathfrak{s}$ as,
\begin{align}\label{eq:gam-analyt}
\gamma_\mathfrak{s} \propto g_X^4\,
\begin{cases}
T^4\,, & T \gg \Mzp/2\,, \\[8pt]
\Mzp^4\,\left(T/\Gamma_{Z'}\right)\,K_1\!\left[\Mzp/T\right]\,, & T \simeq \Mzp/2\,, \\[8pt]
T^8/\Mzp^4\,, & T \ll \Mzp/2\,,
\end{cases}
\end{align}
with corresponding asymptotic DM yield
\begin{align}\label{eq:Y0}
Y_0\propto g_X^4\,
\begin{cases}
M_P/T_0\,, & T \gg \Mzp/2\,, \\[8pt]
M_P/\Gamma_{Z'}\,, & T \simeq \Mzp/2\,, \\[8pt]
M_P\,T_{\rm rh}^3/\Mzp^4\,, & T \ll \Mzp/2\,,
\end{cases}
\end{align}
where $T_{\rm rh}$ corresponds to the maximum temperature of the Universe, considering instantaneous reheating. From Eq.~\eqref{eq:Y0} it is clear that for $T \gg \Mzp/2$, the mediator mass becomes negligible, while for $T \ll \Mzp/2$, the production is suppressed since the mediator is too heavy. Near the resonance, $T \simeq \Mzp/2$, one may apply the narrow-width approximation, whereas far from the resonance, the propagator can be approximated as $1/(s - \Mzp^2)^2$ by neglecting the decay width of the mediator.} The observed DM yield requires to satisfy $Y_0\,m_\chi = \Omega_{\rm DM}\,h^2 \,\frac{\rho_c}{s_0 h^2} \simeq 4.3 \times 10^{-10}\,\text{GeV}$,
where $Y_0 = Y_{\rm DM}(T_0)$ is the present yield, $\rho_c \simeq 1.05 \times 10^{-5}\,h^2\,\text{GeV/cm}^3$ is the critical density, $s_0 \simeq 2.69 \times 10^3\,\text{cm}^{-3}$ is the present entropy density~\cite{ParticleDataGroup:2022pth}, and $\Omega_{\rm DM} h^2 \simeq 0.12$ is the observed value~\cite{Planck:2018vyg}. 

Finally, for freeze-in to be valid, the DM production rate must remain sub-Hubble until freeze-in completes. We check this by comparing the scattering rate $\gamma_{\mathfrak{s}}/n_{\rm eq}^{\rm DM}$ with $H$, where the equilibrium DM number density is
\begin{equation}
n_{\rm eq}^{\rm DM} = \frac{2T}{\pi^2}\,m_\chi^2\,K_2\!\left(\frac{m_\chi}{T}\right)\,.
\end{equation}
From this requirement we derive an upper bound on the coupling as,
\begin{align}
& g_X < 
\begin{cases}
3.1\times10^{-5}\sqrt{\frac{\Mzp}{1\,\text{TeV}}}\,\left(\frac{1\,\text{TeV}}{T}\right)^{1/4}\,, & T\gg\Mzp/2\,,
\\[10pt]
8.3\times 10^{-6}\,\left(\frac{\Mzp}{1\,\text{TeV}}\right)^{3/2}\,\left(\frac{1\,\text{TeV}}{T}\right)^{5/4} \,, & T\ll\Mzp/2\,,
\end{cases}
\end{align}
where $x_H=0,\,x_\Phi=1$ and $n_\chi=100$ are considered. This estimation is conservative since it assumes an equilibrium number density for the DM. 

{\color{black} For $n_\chi=10^4$, the bound becomes considerably more restrictive, particularly for lighter DM masses $\mdm \lesssim 10~\text{GeV}$, as one can see from the top panel of Fig.~\ref{fig1}. This behaviour can be understood from the fact that the DM production cross-section scales as $n_\chi^2$, such that a large DM charge significantly enhances the reaction density. Furthermore, lighter DM experiences weaker phase-space suppression, resulting in more efficient production, especially around the resonant regime $T \sim \Mzp/2$. Consequently, the interaction rate can become comparable to or even exceed the Hubble expansion rate, potentially driving the DM towards thermal equilibrium. However, as stated before, since we are already making a conservative estimation by considering equilibrium DM number density, and therefore adopt $g_X < 10^{-5}$ to ensure out if equilibrium production. For completeness, we also show the $n_\chi=100$ case in the bottom panel.
}
\\ 

(ii) {\it GWs from cosmic strings:} The dominant channel of energy loss from cosmic strings (CS) arises from GW emission by oscillating loops, as demonstrated in simulations based on the Nambu--Goto action~\cite{Ringeval:2005kr,Blanco-Pillado:2011egf}. Here we will closely follow the analysis in~\cite{Barman:2025bir} to estimate bounds on $v_\Phi$, and consequently on $g_X,\,\Mzp$. In this case, the radiated power takes the form~\cite{Vilenkin:1981bx}, 
\begin{equation}
    P_{\rm GW} = \frac{G}{5} (\dddot{Q})^2 \propto G\mu^2\,,
\end{equation}
where $\mu$ denotes the string tension. Consequently, the energy loss rate is  
\begin{equation}
    \frac{dE}{dt} = -\Gamma G\mu^2\,,
\end{equation}
with $\Gamma \simeq 50$~\cite{Vachaspati:1984gt}. 
The length of a loop, initially $l_i = \alpha t_i$, evolves as  
\begin{equation}
    l(t) = \alpha t_i - \Gamma G\mu(t - t_i)\,,
\end{equation}
where $\alpha \sim 0.1$ characterizes the typical loop size~\cite{Blanco-Pillado:2013qja,Blanco-Pillado:2017oxo}. 
Loops emit GWs in discrete harmonics with frequencies  
\begin{equation}
    f_k = \frac{2k}{l(t)}\,, \quad k = 1,2,3,\dots\,.
\end{equation}

The present-day GW spectrum is defined as  
\begin{equation}
    \Omega_{\rm GW}(t_0,f) = \sum_k \Omega_{\rm GW}^{(k)}(t_0,f) 
    = \frac{f}{\rho_c} \frac{d\rho_{\rm GW}}{df}\,,
\label{eqn:omgcs1}
\end{equation}
where the GW energy density redshifts as $a^{-4}$. 
The contribution from each harmonic is given by~\cite{Blanco-Pillado:2013qja}  
\begin{equation}
    \frac{d\rho_{\rm GW}^{(k)}}{df} 
    = \int_{t_F}^{t_0} \left[\frac{a(t_E)}{a(t_0)}\right]^4 
    P_{\rm GW}(t_E,f_k) \frac{dF}{df}\, dt_E\,,
\label{eqn:omgcs2}
\end{equation}
with $dF/df = f[a(t_0)/a(t_E)]$. 
The power radiated per frequency reads  
\begin{equation}
    P_{\rm GW}(t_E,f_k) 
    = \frac{2k G\mu^2 \Gamma_k}{f[a(t_0)/a(t_E)]^2}\,
    n\!\left(t_E, \frac{2k}{f}\left[\frac{a(t_E)}{a(t_0)}\right]\right)\,,
\label{eqn:omgcs3}
\end{equation}
where  
\begin{equation}
    \Gamma_k = \frac{\Gamma\,k^{-4/3}}{\sum_{m=1}^\infty m^{-4/3}}\,, 
    \qquad \sum_k \Gamma_k = \Gamma\,.
\end{equation}
The loop number density $n(t_E,l)$, which depends on the cosmological background $a(t) \propto t^\beta$, is~\cite{Martins:1996jp,Martins:2000cs,Auclair:2019wcv}  
\begin{equation}
    n(t_E,l) = \frac{A_\beta}{\alpha} 
    \frac{(\alpha + \Gamma G\mu)^{3(1-\beta)}}
         {[l + \Gamma G\mu t_E]^{4 - 3\beta} t_E^{3\beta}}\,,
\label{eqn:omgcs4}
\end{equation}
with $A_\beta$ a constant. 
Assuming cusp-dominated emission~\cite{Damour:2001bk,Gouttenoire:2019kij}, the present-day GW spectrum becomes  
\begin{equation}
    \Omega_{\rm GW}^{(k)}(t_0,f) 
    = \frac{2k G\mu^2 \Gamma_k}{f \rho_c} 
    \int_{t_{\rm osc}}^{t_0} dt 
    \left[\frac{a(t)}{a(t_0)}\right]^5 n(t,l_k)\,,
\label{eqn:omgcsfin}
\end{equation}
where $t_{\rm osc}$ marks the end of the friction-dominated epoch~\cite{Vilenkin:1991zk}. 
During the radiation era, the spectrum exhibits a flat plateau with amplitude, 
\begin{equation}\label{eq:GW-plat}
    \Omega_{\rm GW}^{(k=1),{\rm plateau}}(f) 
    = \frac{128\,\pi G\mu}{9\,\zeta(4/3)} \frac{A_r}{\epsilon_r} \Omega_r 
    \left[(1 + \epsilon_r)^{3/2} - 1\right]\,,
\end{equation}
where $\epsilon_r = \alpha / (\Gamma G\mu)$, and $A_r = 0.54$ for radiation domination~\cite{Auclair:2019wcv}. CMB measurements require $G\mu\lesssim  10^{-7}$~\cite{Charnock:2016nzm}, while recent results from NANOGrav~\cite{NANOGrav:2023hvm} puts a strong upper bound: $G\mu\lesssim 10^{-10}$.
\\

\begin{table*}[!t]
\centering
\small
\renewcommand{\arraystretch}{1.2}
\setlength{\tabcolsep}{8pt}
\begin{tabular}{c c c}
\hline\hline
$\Mzp$ [GeV] &
$E_\nu \simeq 3\,\Mzp$ [GeV] &
Relevant experiments \\
\hline
$1-10^2$ & $3-3\times10^2$
& IceCube, ANTARES, Baikal-GVD
\\
$10^2-10^5$ & $3\times10^2-3\times10^5$ & IceCube, ANTARES, Baikal-GVD
\\
$10^5-10^6$ & $3\times10^5-3\times10^6$ & IceCube, Baikal-GVD, KM3NeT
\\
$10^6-10^7$ & $3\times10^6-3\times10^7$ & IceCube Gen2, PAO, KM3NeT
\\
\hline\hline
\end{tabular}
\caption{Correspondence between the $Z^\prime$ mass and the monochromatic neutrino energy predicted in our benchmark scenario $m_\chi=3M_{Z^\prime}$. The last column indicates the experiments with sensitivity in the corresponding neutrino-energy range.}
\label{tab:nu-expt}
\end{table*}
\section{Results and discussions} 
\begin{figure*}
\centering    
\includegraphics[width=0.497\textwidth,angle=0]{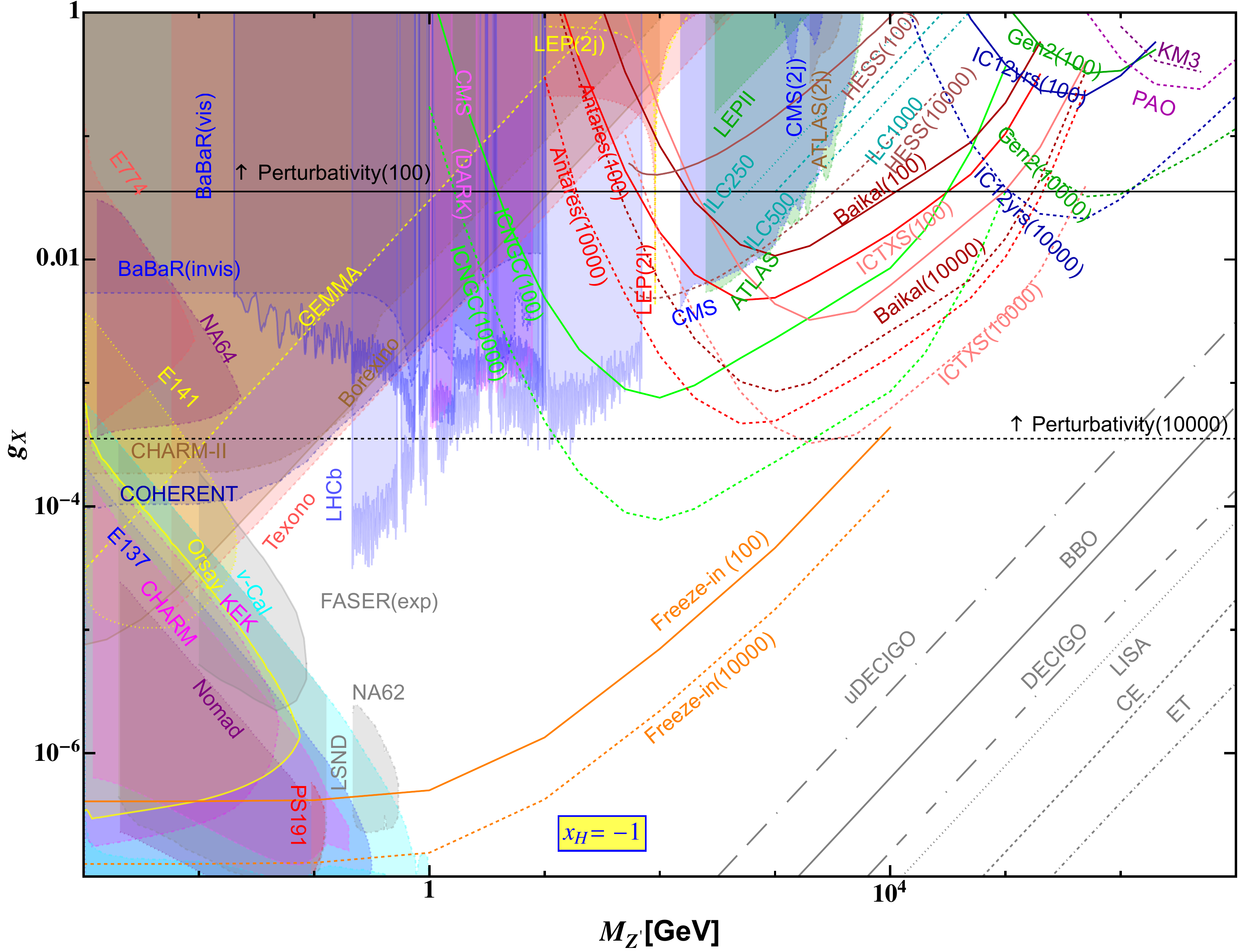}
\includegraphics[width=0.497\textwidth,angle=0]{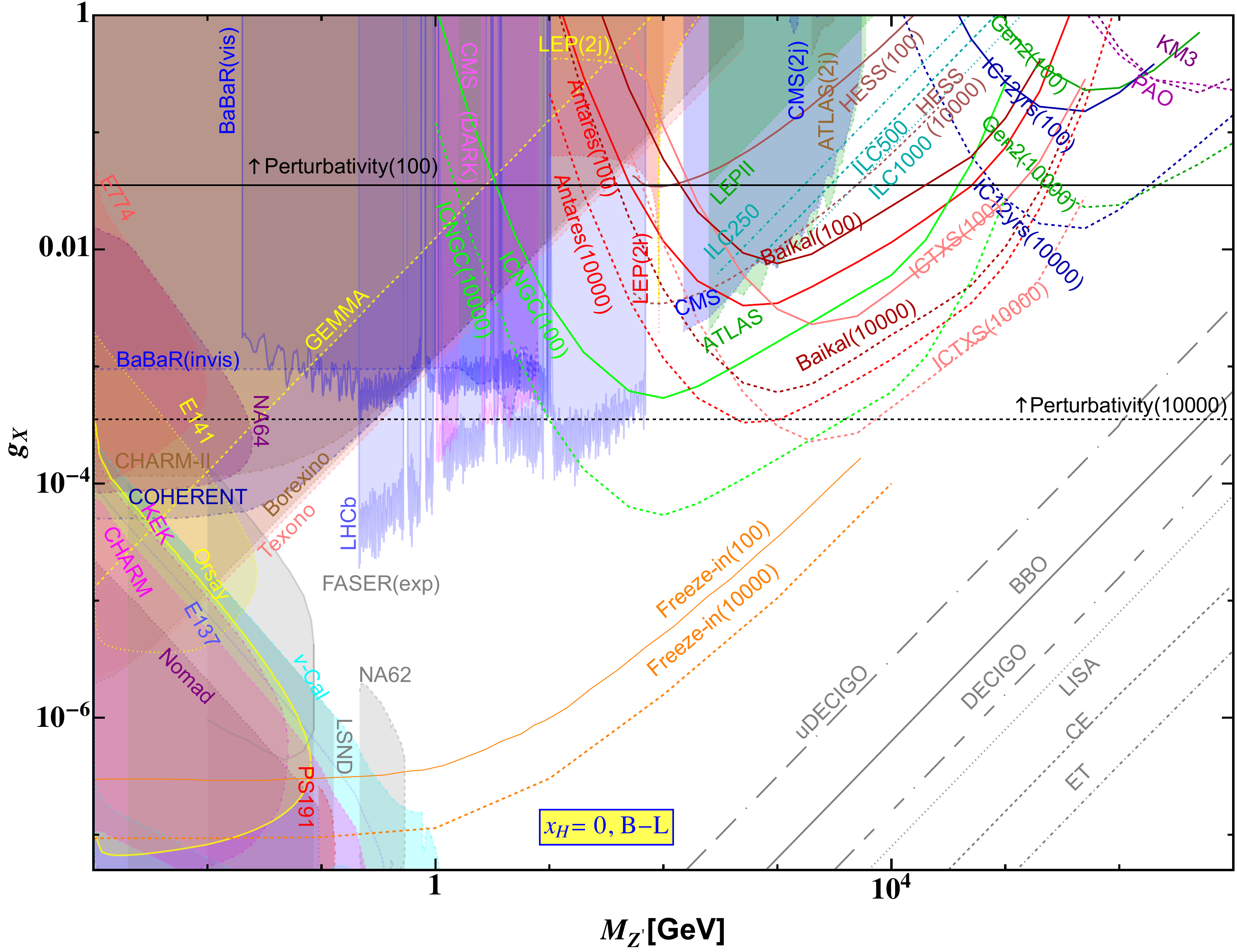}
\includegraphics[width=0.497\textwidth,angle=0]{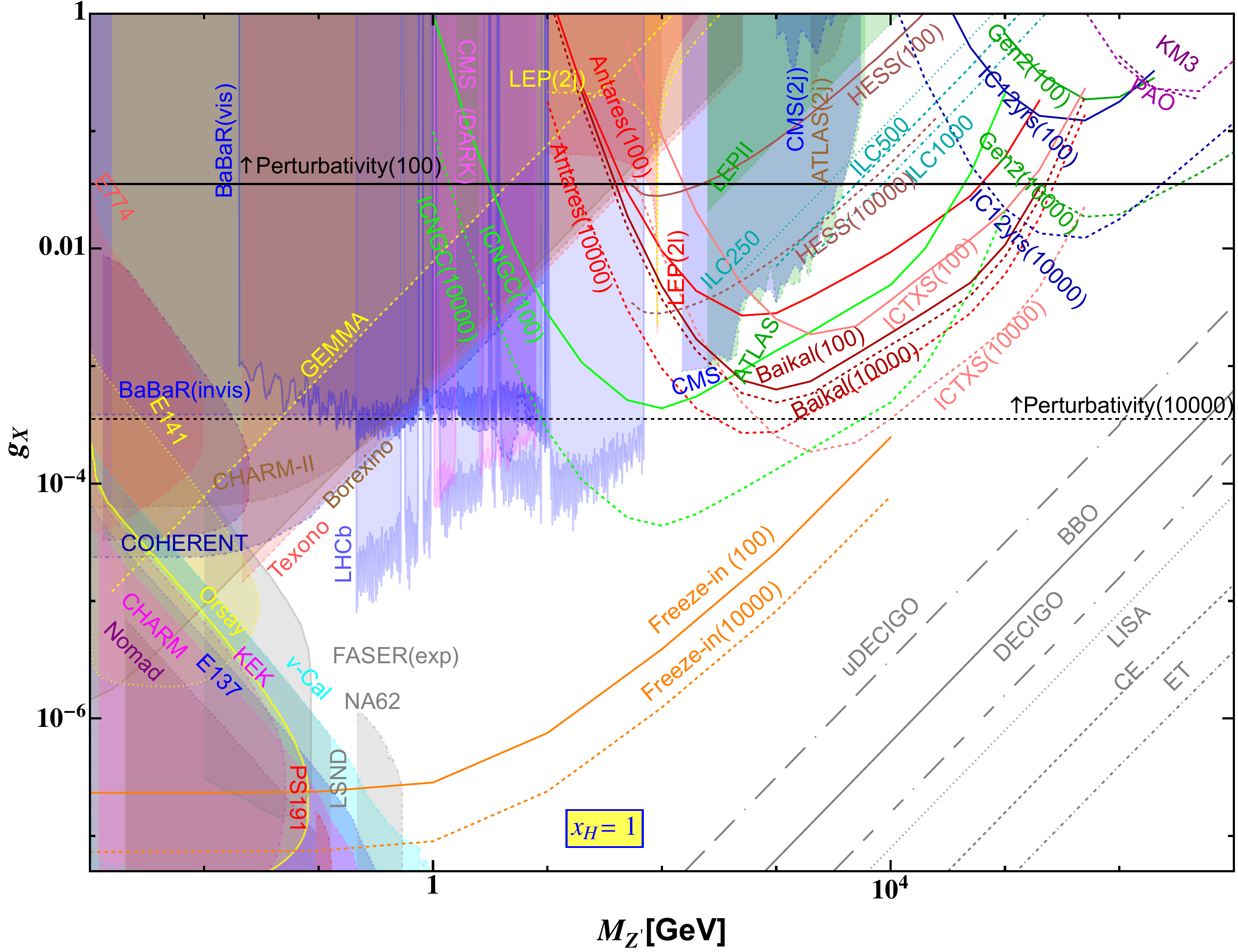}
\includegraphics[width=0.497\textwidth,angle=0]{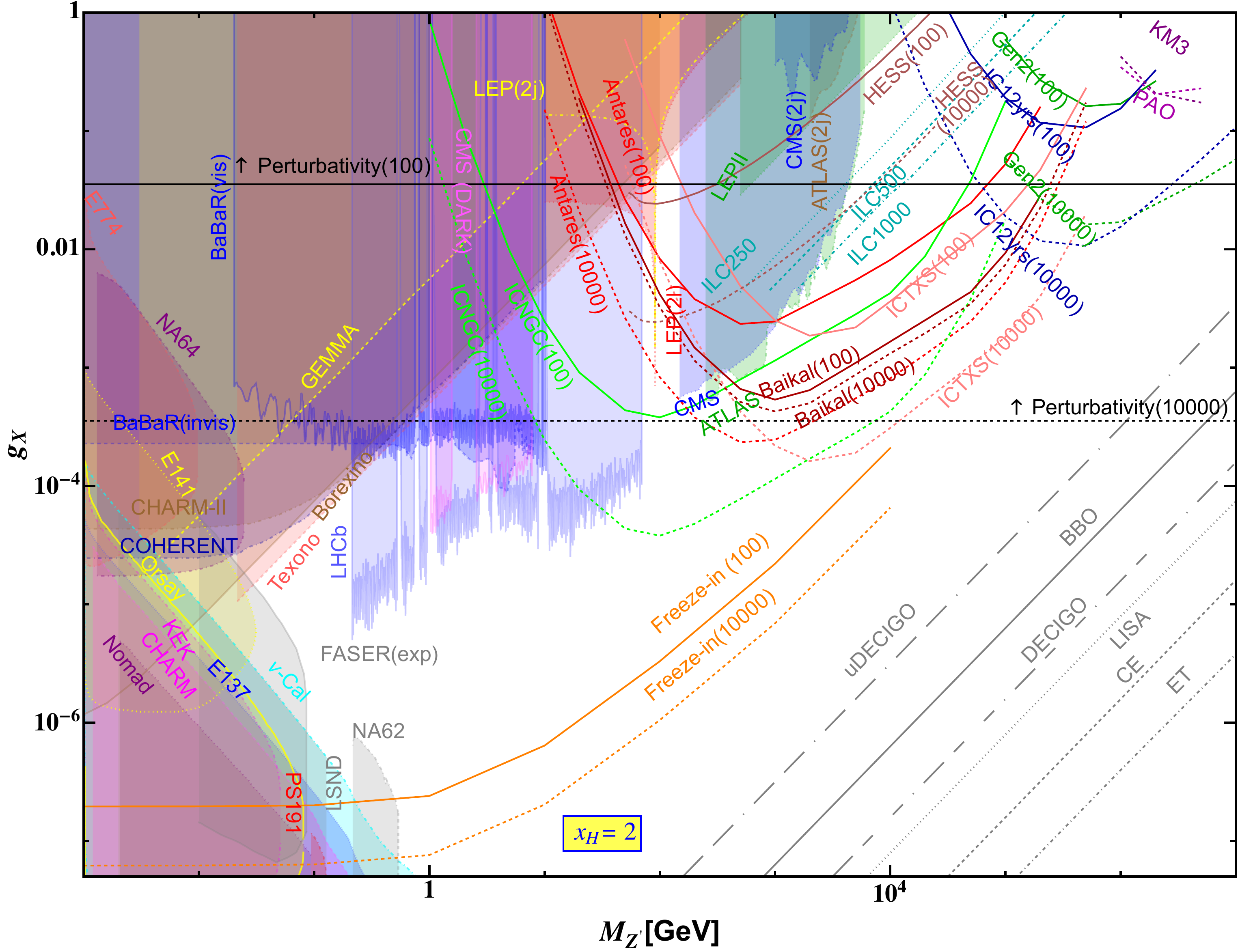}
\caption{Limits on $g_X-M_{Z^\prime}$ plane from DM annihilation into energetic neutrino events observed by IceCube (IC) from AGN NGC1068 (ICNGC) \cite{IceCube:2022der}, cosmic blazar TXS0506+056 (ICTXS) \cite{IceCube:2022der} at IceCube, ANTARES \cite{ANTARES:2024ihw,KM3NeT:2025npi}, Baikal-GVD (Baikal) \cite{Baikal-GVD:2022fis,KM3NeT:2025npi}, IceCube 12.6 years sensitivity (IC12 yrs) \cite{Meier:2024flg}, prospective IceCube Gen2 (Gen2) \cite{IceCube-Gen2:2020qha,Meier:2024flg} for $n_\chi=100$ (solid), $10000$ (dashed) for different $U(1)_X$ charges like $x_H=-1,0$ (B$-$L), $1$ and $2$ considering $x_\Phi=1$. Limits from KM3NeT (KM3) \cite{KM3NeT:2025npi}, HESS \cite{HESS:2022ygk}, and PAO \cite{PierreAuger:2022atd,KM3NeT:2025npi} for $n_\chi=10000$ (dashed) are also shown on the same plane. We compare these results with DM freeze-in via $Z'$ shown by orange solid (dashed) curve for $n_\chi=100$ (10000) and future GW (from cosmic string) search experiments (uDECIGO, BBO, DECIGO, LISA, CE, ET) by gray diagonal contours. Limits from perturbativity for $n_\chi=100$ (10000) are shown by horizonal black solid (dashed) contour considering $n_\chi\, g_X<\sqrt{4 \pi}$. Existing (prospective) limits from different beam-dump (electron, proton), low and high energy scattering experiments are represented by various colored regions.}
\label{fig2}
\end{figure*}
Bounds obtained from energetic cosmic showers observed at {\color{black} IceCube (IceCube-170922A)}, ANTARES, Baikal-GVD, KM3NeT and PAO are shown in Fig.~\ref{fig2} for  $x_H=\{-1,\, 0,\,1,\, 2\}$, considering different DM charges: $n_\chi=100$ (red, pink, green, darker red solid lines) and $n_\chi=10000$ (red, pink, green, darker red dashed lines), taking into account the best-fit values of each dataset. Events from IceCube (NGC 1068 and TXS0506+056){\footnote{{\color{black}We compute the neutrino flux using the DM density profile calibrated for NGC 1068 as a representative AGN benchmark. Although NGC 1068 and TXS 0506+056 differ in several astrophysical properties (e.g., redshift, black hole mass, and host environment), both belong to the AGN class and are independently reported neutrino source candidates in IceCube analyses. In this work, we do not assume identical source properties; rather, we use the model predictions as a common theoretical framework to derive constraints on the DM model parameters by comparing with the observed IceCube event information from both NGC 1068 and TXS 0506+056.}}}, ANTARES, Baikal-GVD (Baikal) could provide stronger bounds on $g_X$ compared to the existing bounds obtained by the CMS and ATLAS detectors, considering dilepton and dijet final states from TeV scale $Z^\prime$. These bounds can be stronger than the observed LEP bounds~\cite{Asai:2023mzl} at the $Z-$pole comparing dilepton and dijet final states, LEP-II bounds and prospective sensitivities from the International Linear Collider (ILC) obtained by analyzing effective vertex scenario in \cite{Das:2021esm}. We find that bounds obtained on $g_X$ for $M_{Z^\prime} < 70$ GeV are weaker than those obtained from the dark photon search experiments at the LHCb~\cite{LHCb:2019vmc}, CMS~\cite{CMS:2023hwl} and BaBaR~\cite{BaBar:2014zli,BaBar:2017tiz}. {\color{black} It is important to highlight that possible tension between the KM3NeT single-event observation and IceCube’s lack of observation of any event exceeding 10 PeV is well-known, and several studies have attempted to address this~\cite{Brdar:2025azm,Farzan:2025ydi,Dev:2025czz,Li:2025tqf}. Our intention here is however not to derive a statistically rigorous exclusion bound from the single KM3NeT event, but rather to use it as an illustrative benchmark for the flux level required by a monochromatic neutrino signal from DM annihilation. We assume that the observed event can be saturated by the DM-induced neutrino flux and derive the corresponding parameter-space requirement. Given that the inference is based on a single event, the associated flux normalization is subject to significant uncertainty and should not be interpreted as a robust exclusion limit.}

Strong bounds are obtained by the neutrino-electron scattering experiments from BOREXINO~\cite{Borexino:2000uvj,Bellini:2011rx}, TEXONO~\cite{TEXONO:2009knm}, GEMMA~\cite{Beda:2010hk}, CHARM-II~\cite{CHARM-II:1993phx, CHARM-II:1994dzw} and neutrino-nucleon scattering from COHERENT~\cite{COHERENT:2018imc,COHERENT:2020iec,COHERENT:2020ybo} experiment. These mostly affect light $Z^\prime$ having mass lighter than $\mathcal{O}(1)$ GeV.  We compare our bounds with the limits obtained by perturbativity $(n_\chi g_X^{} < \sqrt{4 \pi})$ considering $n_\chi=100$ (black solid) and 1000 (black dashed). Constraints from KM3NET (KM3), PAO and prospective IceCube 12 years data (Gen2) will be extremely weak compared to the respective perturbativity bounds within $10^{5}$ GeV $\leq M_{Z^\prime} \leq 10^{7}$ GeV. We find, limits from IceCube (NGC 1068 and TXS0506+056), ANTARES, Baikal-GVD for $n_\chi=100$ (red, pink, green, darker red solid lines) are stronger than the corresponding perturbativity bound (black solid) within $100$ GeV $\leq M_{Z^\prime} \leq 10^{5}$ GeV depending on $x_H$. For $n_\chi=10000$ (black dashed) we see corresponding bounds from IceCube (NGC 1068 and TXS0506+056) and ANTARES (red, green dashed lines) are stronger than perturbative bounds within $10$ GeV $\leq M_{Z^\prime} \leq 10^{4}$ GeV, whereas bounds from Baikal are slightly weaker. Depending on the observed neutrino flux, $x_H$, $n_\chi$ and perturbativity constrains, strongest bound on $g_X$ lies in the range $10^{-4}\lesssim g_X^{}\lesssim 10^{-3}$, for $M_{Z^\prime}\simeq\mathcal{O}(1)$ TeV. {\color{black} Note that, on our setup, due to the the benchmark choice $\mdm=3\,\Mzp$, monochromatic neutrino energy produced from DM annihilation is fixed by $E_\nu\simeq\mdm=3\,\Mzp$. Therefore, the neutrino energy spans a wide range depending on the $Z'$ mass.  In Tab.~\ref{tab:nu-expt} we have provided a mapping between the model parameters: $[\mdm,\,\Mzp]$, and the relevant experiments capable of probing corresponding energy range.}

Recently, the H.E.S.S.\ Collaboration reported a new search for DM annihilation in the central region of the Milky Way halo using an unprecedented very-high-energy dataset $(\gtrsim 100\,\mathrm{GeV})$ collected with the five-telescope H.E.S.S.\ array targeting the Galactic Centre. Based on observations from 2014 to 2020, constraints were derived on the velocity-averaged annihilation cross sections,
{\color{black}
\begin{align}
& \langle\sigma v\rangle_{\tau^+\tau^-}= \frac{n_\chi^2 g_X^4 }{ 2\pi m_\chi^2 }\,\frac{\big(1-m_{\tau}^2/m_{\chi}^2\big)^{3/2}\,(x_\Phi^{}+x_H/2)^2}{(x_{z^\prime}^2-4)^2 + x_{z^\prime}^4\Gamma_{Z^\prime}^2/M_{Z^\prime}^2}\,,
\end{align}
}
considering \(\tau^+\tau^-\) channel. For \(x_H=-1\), we find,
\(6\times10^{-2}\lesssim g_X \lesssim 2.3\), corresponding to a \(Z'\) mass in the range
\(540\,\mathrm{GeV}\lesssim M_{Z'} \lesssim 2.6\times10^{4}\,\mathrm{GeV}\) for
\(n_\chi=100\). The constraint tightens for \(n_\chi=10^{4}\),
yielding \(5\times10^{-3}\lesssim g_X \lesssim 0.2\). The orange solid curves in Fig.~\ref{fig2} correspond to contours of right DM abundance.  We compare our results with the existing bounds~\cite{Asai:2022zxw} on the chiral scenarios from
proton beam-dump involving Nomad
\cite{NOMAD:2001eyx}, $\nu-$cal \cite{Blumlein:1990ay,Barabash:2002zd}, CHARM \cite{CHARM:1985nku}, FASER \cite{FASER:2023tle,FASER:2024bbl}, NA62 \cite{NA62:2025yzs}, LSND \cite{LSND:1997vqj,LSND:2001aii}, PS191 \cite{Bernardi:1985ny}
experiments and electron beam-dump involving E141 \cite{Riordan:1987aw}, E137 \cite{Bjorken:1988as}, E774 \cite{Bross:1989mp}, Orsay \cite{Davier:1989wz}, KEK \cite{Konaka:1986cb} experiments respectively. We find that parameter space $M_{Z^\prime} \leq 1$ GeV is tightly constrained by these results, and freeze-in provides stringent bound within $1$ GeV $\leq M_{Z^\prime} \leq 10^{4}$ GeV, demanding $7\times 10^{-8} \lesssim g_X < 10^{-4}$ depending on $x_H$ and $n_\chi$. These bounds are much stronger than those obtained from the scattering experiments and perturbativity. 

From Eq.~\eqref{eq:GW-plat}, note that, $\Omega_{\rm GW}^{(k=1)}(f)\propto v_{\Phi}$, and hence a higher symmetry breaking scale is more likely to be probed by the GW detectors. The diagonal gray lines (solid and broken) denote a few such benchmark $v_{\Phi}$'s that are within the reach of several futuristic GW detectors, such as, Big Bang Observer (BBO)~\cite{Crowder:2005nr, Corbin:2005ny}, ultimate DECIGO (uDECIGO)~\cite{Seto:2001qf, Kudoh:2005as}, LISA~\cite{LISA:2017pwj}, the cosmic explorer (CE)~\cite{Reitze:2019iox} and the Einstein Telescope (ET)~\cite{Hild:2010id, Punturo:2010zz, Sathyaprakash:2012jk, Maggiore:2019uih}. These prospective experiments could provide strong constraints typically for $\Mzp\gtrsim 1$ TeV.

\section{Conclusions}
We explore a general $U(1)$ extension of the Standard Model where the new gauge boson $Z^\prime$ couples asymmetrically to left- and right-handed fermions, rendering the framework inherently chiral. To confront this scenario with observations, we consider the high energy neutrino flux measured at IceCube, ANTARES, Baikal-GVD, KM3NeT and PAO, assuming its origin lies in dark matter annihilation within astrophysical environments such as active galactic nuclei and blazars. {\color{black} Our approach is purely phenomenological, based on flux comparisons rather than event-level statistical analyses, under the conservative assumption that the observed events of a given experiment are fully saturated by dark matter annihilation.} Adopting a Dirac-type dark matter candidate, we derive constraints in the $[g_X, M_{Z^\prime}]$ parameter space for different $U(1)$ charge assignments. We further examine bounds arising from the relic abundance of dark matter, as measured by Planck, within a freeze-in production mechanism. Our results are compared with current and projected limits from both low- and high-energy scattering experiments, as well as from electron and proton beam-dump searches. Furthermore, gravitational waves sourced by cosmic strings may provide complementary constraints in the regime of a heavy $Z^\prime$. We find that bounds from neutrino telescopes can be comparable (or even stronger) to those from colliders, particularly for large $M_{Z^\prime}$. Although freeze-in contours generally evade the reach of neutrino telescopes while lying within the sensitivity of beam-dump experiments for $M_{Z^\prime} \lesssim 1$ GeV. Future experimental improvements could significantly tighten the viable parameter space for dark matter.

\vspace{-0.098in}
\bibliographystyle{utphys}
\bibliography{bibfile}
\end{document}